\documentclass[11pt]{article}

\usepackage{hyperref}
\usepackage{verbatim}
\usepackage{graphics}
\usepackage{graphicx}
\usepackage{bbm}
\usepackage{caption}
\usepackage{subcaption}
\usepackage[round,authoryear,comma]{natbib}
\usepackage[noadjust]{cite}
\usepackage{parskip}
\usepackage{mathrsfs}
\usepackage{fullpage}
\usepackage{amsfonts,latexsym,eucal,amsmath,amsthm,amssymb,bm,pdfpages}
\usepackage{array}
\usepackage[T1]{fontenc}
\usepackage[ansinew]{inputenc}
\usepackage{enumerate}
\usepackage{booktabs}
\usepackage{algorithm,algorithmic}
\usepackage{rotfloat}
\usepackage{booktabs}
\usepackage{multirow}
\usepackage{listings}

\renewcommand{\tilde}{\widetilde}
\renewcommand{\hat}{\widehat}
\renewcommand{\Pr}{\mathbb{P}}
\renewcommand{\Re}{\mathbb{R}}
\newcommand{\Ex}{\mathbb{E}}

\newcommand{\mX}{\mathcal{X}}

\newtheorem{theorem}{Theorem}

\newtheorem{corollary}[theorem]{Corollary}
\theoremstyle{definition}
\newtheorem*{remark*}{Remark}
\newtheorem{definition}[theorem]{Definition}

\renewcommand{\Re}{\mathbb{R}}

\newcommand{\hX}{\widehat X}

\graphicspath{{./Figures/}}%

\begin{document}
\title{Empirical Bayes Estimation for the Stochastic Blockmodel}
\author{
Shakira Suwan, Dominic S. Lee\\
Department of Mathematics and Statistics\\
University of Canterbury\\
\vspace*{0.15 in}
Christchurch, New Zealand\\
Runze Tang, Daniel L. Sussman, Minh Tang, Carey E. Priebe\\ 
Department of Applied Mathematics and Statistics\\
Johns Hopkins University\\
Baltimore, Maryland, USA\\
}

\maketitle
\begin{abstract}
\noindent 
Inference for the stochastic blockmodel is currently of burgeoning interest in the statistical community,
as well as in various application domains
 as diverse as social networks, citation networks, brain connectivity networks (connectomics), etc.
Recent theoretical developments have shown that spectral embedding of graphs yields tractable distributional results;
in particular, a random dot product latent position graph formulation of the stochastic blockmodel
informs a mixture of normal distributions for the adjacency spectral embedding.
We employ this new theory to provide an empirical Bayes methodology for estimation of block memberships of vertices
in a random graph drawn from the stochastic blockmodel, and demonstrate its practical utility. 
The posterior inference is conducted using a Metropolis-within-Gibbs algorithm.
The theory and methods are illustrated through Monte Carlo simulation studies,
both within the stochastic blockmodel and beyond, and experimental results on a Wikipedia graph are presented.
\end{abstract}

\section{Introduction}\label{intro}

The stochastic blockmodel (SBM) is a generative model for network data introduced in 
\citet{holland1983stochastic}.
The SBM is a member of the general class of latent position random graph models introduced in 
\citet{hoff2002}.
These models have been used in various application domains as diverse as
social networks (vertices may represent people with edges indicating social interaction),
citation networks (who cites whom),
connectomics (brain connectivity networks;
   vertices may represent neurons with edges indicating axon-synapse-dendrite connections, or
   vertices may represent brain regions with edges indicating connectivity between regions),
and many others.
For comprehensive reviews of statistical models and applications, 
see \citet{fienberg2010introduction,goldenberg2010survey,fienberg2012brief}.
In general, statistical inference on graphs is becoming essential in many areas of science, engineering, and business.

The SBM supposes that each of $n$ vertices is assigned to one of $K$ blocks. The probability of an edge between two vertices depends only on their respective block memberships, and the presence of edges are conditionally independent given block memberships. By letting $\tau_i$ denote the block to which vertex \textit{i} is assigned, a $K \times K$ matrix $B$ is defined as the probability matrix such that the entry $B_{\tau_i,\tau_j}$ is the probability of an edge between vertices $i$ and $j$. The block proportions are represented by a $K$-dimensional probability vector $\rho$. 
Given an SBM graph, estimating the block memberships of vertices is an obvious and important task. 
Many approaches have been developed for estimation of vertex block memberships, including 
likelihood maximization~\citep{bickel2009,choi2012,celisse2012,bickel2013},
maximization of modularity~\citep{newman2006}, 
spectral techniques~\citep{rohe2011spectral,sussman2012consistent,fishkind2013consistent}, and 
Bayesian methods~\citep{snijders1997,nowicki2001,handcock2007,airoldi2008}.

Latent position models for random graphs provide a framework in which graph structure is parametrized by a latent vector associated with each vertex.
In particular, this paper considers the special case of the latent position model known as the random dot product graph model (RDPG),
 introduced in~\citet{nickel2006random} and~\citet{young2007random}.
In the RDPG, each vertex is associated with a latent vector,
and the presence or absence of edges are independent Bernoulli random variables, conditional on these latent vectors. 
The probability of an edge between two vertices is given by the dot product of the corresponding latent vectors.
An SBM can be defined in terms of an RDPG model for which all vertices that belong to the same block share a common latent vector.

When analyzing RDPGs, 
the first step is often to estimate the latent positions, and these estimated latent positions can then be used for subsequent analysis. 
Obtaining accurate estimates of the latent positions will consequently give rise to accurate inference ~\citep{sussman2014consistent}, as the latent vectors determine the distribution of the random graph. 

\citet{sussman2012consistent} describes a method for estimating the latent positions in an RDPG 
using a truncated eigen-decomposition of the adjacency matrix.

\citet{athreya2013limit} proves that for an RDPG, 
the latent positions estimated using adjacency spectral graph embedding converge in distribution to a multivariate Gaussian mixture. 
This suggests that we may consider the estimated latent positions of a $K$-block SBM
as (approximately) an independent and identically distributed sample from a mixture of $K$ multivariate Gaussians. 

In this paper, we demonstrate the utility of an estimate of this multivariate Gaussian mixture as an empirical prior distribution in a Bayesian inference methodology for estimating block memberships in an SBM graph, as it quantifies residual uncertainty in the model parameters after adjacency spectral embedding.

This paper is organized as follows. In Section~\ref{section:EB}, we formally present the SBM as an RDPG model 
and describe how the theorem of~\citet{athreya2013limit} motivates our mixture of Gaussians empirical prior. 
We then present our empirical Bayes methodology for estimating block memberships in the SBM,
and the Markov chain Monte Carlo (MCMC) algorithm that implements the Bayesian solution. 
In Section~\ref{section:eg}, we present simulation studies and an experimental analysis demonstrating the performance of our empirical Bayes methodology.
Finally, Section~\ref{section:discussion} discusses further extensions and provides a concluding summary.

\section{Background}\label{section:EB}

Network data on $n$ vertices may be represented as an adjacency matrix $A \in \{0,1\}^{n \times n}$.
We consider simple graphs, so that $A$ is 
symmetric (undirected edges imply $A_{ij} = A_{ji}$ for all $i,j$),
hollow (no self-loops implies $A_{ii} = 0$ for all $i$), and
binary (no multi-edges or weights implies $A_{ij} \in \{0,1\}$ for all $i,j$).
For our random graphs, the vertex set is fixed; it is the edge set that is random.


Let $\mX \subset \mathbb{R}^d$ be a set such that $x,y \in \mX$ implies $\langle x,y \rangle \in [0,1]$,
let $X_i \stackrel{iid}{\sim} F$ on $\mX$,
and write $X=[X_1|\dotsc|X_n]^\top\in \mathbb{R}^{n \times d}$.
\\

\begin{definition}[Random Dot Product Graph]\label{def:rdpg}
A random graph $G$ with adjacency matrix $A$
is said to be a random dot product graph (RDPG) if
\[ \Pr[A| X] = \prod_{i < j}  \langle X_i,X_j \rangle^{A_{ij}}(1-\langle X_i,X_j \rangle)^{1-A_{ij}}. \]
\end{definition}

Thus, in the RDPG model, each vertex $i$ is associated with a latent vector $X_i$.
Furthermore, conditioned on the latent positions $X$,
the edges $A_{ij} \stackrel{ind}{\sim}\mathrm{Bern}(\langle X_i, X_j \rangle)$.

For an RDPG, we also define the $n \times n$ edge probability matrix $P=XX^T$;
 $P$ is symmetric and positive semidefinite and has rank at most $d$.
Hence, $P$ has a spectral decomposition given by  $P=[U_P|\widetilde U_P][S_P\oplus\widetilde S_P][U_P|\widetilde U_P]^T$ where $[U_P|\widetilde U_P]\in \Re^{n\times n}$, and $U_P\in \Re^{n\times d}$ has orthonormal columns, and $S_P\in \Re^{d\times d}$ is diagonal matrix with non-negative non-increasing entries along the diagonal. It follows that there exists an orthonormal $W_n \in \Re^{d \times d}$ such that $U_PS_P^{1/2}=XW_n$. This introduces obvious non-identifiability since $XW_n$ generates the same distribution over adjacency matrices (i.e. $(XW_n)*(XW_n)^\top = XX^\top$). As such, without loss of generality, we consider \textit{uncentered} principal components (UPCA) of $X$, $\tilde{X}$, such that $\tilde X=U_p S_P^{1/2}$. Letting $U_A\in \Re^{n\times d}$ and $S_A \in \Re^{d\times d}$ be the adjacency matrix versions of $U_P$ and $S_P$, the adjacency spectral graph embedding (ASGE) of $A$ to dimension $d$ is given by  $\hX=U_A S_A^{1/2}$.

The SBM can be formally defined as an RDPG for which all vertices that belong to the same block share a common latent vector, according to the following definition.\\

\begin{definition}[(Positive Semidefinite) Stochastic Blockmodel] 
An RDPG can be parameterized as an SBM with $K$ blocks if the number of distinct rows in $X$ is $K$.  
That is, let the probability mass function $f$ associated with the distribution $F$ of the latent positions $X_i$ 
be given by the mixture of point masses $f = \sum_k \rho_k \delta_{\nu_k}$,
where the probability vector $\rho \in (0,1)^K$ satisfies $\sum_{k=1}^{K} \rho_k = 1$ 
and the distinct latent positions are represented by $\nu = [\nu_1|\cdots|\nu_K]^\top \in \mathbb{R}^{K \times d}$.
Thus the standard definition of the SBM with parameters $\rho$ and $B = \nu\nu^\top$
is seen to be an RDPG with $X_i \stackrel{iid}{\sim} \sum_k \rho_k \delta_{\nu_k}$.
\label{def:SBM}\end{definition}
 
Additionally, for identifiability purposes, we impose the constraint that the block probability matrix $B$ have distinct rows;
that is, $B_{k,\cdot} \neq B_{k',\cdot}$ for all $k \neq k'$.

In this setting, the block memberships $\tau_1,\dotsc,\tau_n |K,\rho \overset{iid}{\sim}$ Discrete$([K],\rho)$ such that $\tau_i=\tau_j$ if and only if $X_i=X_j$.
Let $N_k$ be the number of vertices such that $\tau_i=k$; we will condition on $N_k = n_k$ throughout.
Given a graph generated according to the SBM, our goal is to assign vertices to their correct blocks. 


To date, Bayesian approaches for estimating block memberships in the SBM
have typically involved a specification of the prior on the block probability matrix $B=\nu\nu^\top$;
the beta distribution (which includes the uniform distribution as a special case) is often chosen as the prior~\citep{snijders1997,nowicki2001}. 
Facilitated by our re-casting of the SBM as an RDPG and
motivated by recent theoretical advances described in Section~\ref{ebsbm} below,
we will instead derive an empirical prior for the latent positions $\nu$ themselves.

\section{Model}\label{models}
This section presents the models and algorithms we will use to investigate the utility of the empirical Bayes methodology for estimating block memberships in an SBM graph as detailed in Section~\ref{ebsbm} and referred to as \emph{ASGE}.

For comparison purposes, in Sections~\ref{subsubsection:flat} and~\ref{alternatives} we construct an alternative \emph{Flat} and two benchmark models, as outlined below. Note that all four models are named after their respective prior distributions used for the latent positions $\nu$.
\begin{itemize}
\item \textbf{Flat} -- an alternative to the proposed empirical Bayes prior distribution for $\nu$. Since in the absence of the ASGE theory a natural choice for the prior on $\nu$ is the uniform distribution on the parameter space. 
\item \textbf{Exact} -- a primary benchmark model where all model parameters, except the block membership vector $\tau$, are assumed known.
\item \textbf{Gold} -- a secondary benchmark model where $\nu$ and $\tau$ are the unknown parameters; the gold standard mixture of Gaussians prior distribution for $\nu$ takes its hyperparameters to be the true latent positions and theoretical limiting covariances motivated by the distributional results from~\citet{athreya2013limit} presented in Section~\ref{ebsbm}.
\end{itemize} %

\subsection{The Empirical Bayes with ASGE Prior Model (``ASGE'') }\label{ebsbm}

Recently,~\citet{athreya2013limit} proved that for an RDPG
the latent positions estimated using adjacency spectral graph embedding converge in distribution to a multivariate Gaussian mixture. 
We can express this more formally in a central limit theorem (CLT) for the scaled differences between the estimated and true latent positions of the RDPG graph, as well as a corollary to motivate our empirical Bayes prior (henceforth denoted \textit{ASGE}).
\\

\begin{theorem}[\citet{athreya2013limit}]\
Let $G$ be an RDPG with $d$-dimensional latent positions $X_1,\dotsc,X_n \stackrel{iid}{\sim}F$,
and assume distinct eigenvalues for the second moment matrix of $F$.
Let $\tilde{X}\in\Re^{n\times d}$ be the UPCA of $X$ so that the columns of $\tilde{X}$ are orthogonal, and let $\hat X$ be the estimate for $X$. Let $\mathcal{N}(0, \Sigma) $ represent the cumulative distribution function for the multivariate normal, with mean $0$ and covariance matrix $\Sigma$. 
Then for each row $\tilde{X}_i$ of $\tilde{X}$ and $\hat{X}_i$ of $\hat{X}$, 
\[ \sqrt{n}(\tilde{X}_i-\hat{X}_i) \stackrel{\mathcal{L}}{\to} \int \mathcal{N}(0, \Sigma(x)) dF(x)\]
where the integral denotes a mixture of the covariance matrices and, with the second moment matrix $\Delta=\Ex[X_1X_1^\top]$,
\[ \Sigma(x) =\Delta^{-1}\Ex[X_jX_j^\top (x^\top X_j -(x^\top X_j)^2)]\Delta^{-1}.\]
\label{thm:clt}
\end{theorem}

The special case of the SBM gives rise to the following corollary.
\\

\begin{corollary}
In the setting of Theorem~\ref{thm:clt}, suppose $G$ is an SBM with $K$ blocks.
Then, if we condition on $X_i=\nu_k$, we obtain
\begin{equation}
P\left(\sqrt{n}\left( \hat{X}_i - \nu_k\right)\leq z\bigg| X_i=\nu_k\right) {\to} \Phi(z,\Sigma_k) \label{eq:condConv}
\end{equation}
where $\Sigma_k=\Sigma(\nu_k)$ with $\Sigma(\cdot)$ is as in Theorem~\ref{thm:clt}.
\label{cor:sbmCLT}
\end{corollary}

Note that the distribution $F$ of the latent positions $X$ remains unchanged, as $n \to \infty$. 

This gives rise to the mixture of normals approximation
$\hX_1,\cdots,\hX_n \stackrel{iid}{\sim} \sum_k \rho_k \varphi_k$
for the estimated latent positions obtained from the adjacency spectral embedding.
That is, based on these recent theoretical results, we can consider the estimated latent positions as (approximately) an independent and identically distributed sample from a mixture of multivariate Gaussians. 

A similar Bayesian method for latent position clustering of network data is proposed in~\citet{handcock2007}.
Their latent position cluster model is an extension of~\citet{hoff2002},
wherein all the key features of network data are incorporated simultaneously -- 
namely clustering, 
       transitivity (the probability that the adjacent vertices of a vertex having a connection), and 
       homophily on attributes (the tendency of vertices with similar features to possess a higher probability of presenting an edge). 
The latent position cluster model is similar to our model,
but they use the logistic function instead of the dot product as their link function.

Our theory gives rise to a method for obtaining an empirical prior for $\nu$ using the adjacency spectral embedding. Given the estimated latent positions $\hat{X}_1,\dotsc,\hat{X}_n$ 
obtained via the spectral embedding of the adjacency matrix $A$, the next step is to cluster these $\hat{X}_i$ using Gaussian mixture models (GMM).
There are a wealth of methods available for this task; we employ the model-based clustering of \citet{FRmclust} via the \texttt{R} package \texttt{MCLUST} which implements an Expectation-Maximization (EM) algorithm for maximum likelihood parameter estimation.  
This mixture estimate, 
in the context of Corollary~\ref{cor:sbmCLT}, 
 quantifies our uncertainty about $\nu$, suggesting its role as an empirical Bayes prior distribution.
That is, our empirical Bayes prior distribution for $\nu$ is expressed as
\begin{equation}
f(\nu|\{\hat{\mu}_k\}, \{\hat{\Sigma}_k\}) \propto \mathbb{I}_{\mathcal{S}}(\nu) \prod_{k=1}^K \mathcal{N}_d(\nu_k| \hat{\mu}_k, \hat{\Sigma}_k) \label{eq:ebPrior}
\end{equation}
where $\mathcal{N}_d(\nu_k| \hat{\mu}_k, \hat{\Sigma}_k)$ is the density function of a multivariate normal distribution with mean $\hat{\mu}_k$ and covariance matrix $\hat{\Sigma}_k$ denoting standard maximum likelihood estimates (via Expectation-Maximization algorithm) based on the estimated latent positions $\hat{X}_i$ and the indicator $\mathbb{I}_{\mathcal{S}}(\nu)$ enforces homophily and block identifiability constraints for the SBM via
\[
  \mathcal{S} = \{\nu\in \Re^{K\times d}: 
 0\leq \langle\nu_i,\nu_j \rangle \leq \langle \nu_i, \nu_i \rangle\leq 1 ~ \forall i,j\in[K]
\mbox{~ and ~}
 \langle\nu_i,\nu_i\rangle \geq \langle\nu_j,\nu_j\rangle ~ \forall i > j\}.
\]
Algorithm~\ref{algo:EB} provides steps for obtaining the empirical Bayes prior using the ASGE and GMM.

\begin{algorithm}[htpb]
\caption{Empirical Bayes estimation using the adjacency spectral embedding empirical prior}
\label{algo:EB}
\begin{algorithmic}[1]
        \STATE Given graph $G$
        \STATE Obtain adjacency spectral embedding $\hat{X}$
        \STATE Obtain empirical prior via GMM of $\hat{X}$
        \STATE Sample from the posterior via Metropolis--Hasting--within--Gibbs (see Algorithm~\ref{algo:gibbs} below) 
\end{algorithmic}
\end{algorithm}

In the setting of Corollary~\ref{cor:sbmCLT}, for an adjacency matrix $A$, 
the likelihood for the block membership vector $\tau\in[K]^n$ and the latent positions $\nu \in\Re^{K\times d}$ is given by
\begin{equation}\label{likelihood}
  f(A \mid \tau,\nu) = \prod_{i<j} \langle\nu_{\tau_i},\nu_{\tau_j}\rangle^{A_{ij}} (1-\langle\nu_{\tau_i},\nu_{\tau_j}\rangle)^{1-A_{ij}}.
\end{equation}

This is the case where the block memberships~$\tau$, the latent positions~$\nu$, and the block membership probabilities~$\rho$ are assumed unknown. Thus, our empirical posterior distribution for the unknown quantities is given by
\[
  f(\tau,\nu, \rho\mid A) \propto f(A \mid \tau,\nu) f(\tau\mid \rho) f(\rho\mid \theta) f(\nu\mid \{\hat{\mu}_k\},\{\hat{\Sigma}_k\}),
\]
where a multinomial distribution is posited as a prior distribution on $\tau$ with the hyperparameter $\rho$, chosen to follow a Dirichlet distribution with parameters $\theta_k = 1$ for all $k \in K$ in the unit simplex $\Delta_K$, and a multivariate normal prior on $\nu$ as expressed in Eqn~\ref{eq:ebPrior}. To summarize, the prior distributions on the unknown quantities $\tau$, $\nu$, and $\rho$ are
\[\tau \mid \rho \sim \mathrm{Multinomial}(\rho),\]
\[\rho \sim \mathrm{Dirichlet}(\theta),\]
\[\nu \mid \{\hat{\mu}_k\}, \{\hat{\Sigma}_k\} \sim \mathbb{I}_{\mathcal{S}}(\nu) \prod_{k=1}^K \mathcal{N}_d(\nu_k \mid \hat{\mu}_k, \hat{\Sigma}_k).\]
By choosing a conjugate Dirichlet prior for $\rho$, we can marginalize the posterior distribution over $\rho$ as follows:
\begin{align*}
f(\tau,\nu|A) &= \int_{\Delta_K} f(\tau,\nu,\rho|A)d\rho \\
&\propto f(A|\tau,\nu)f(\nu|\{\hat{\mu}_k\},\{\hat{\Sigma}_k\}) \int_{\Delta_K} f(\tau|\rho) f(\rho|\theta) d\rho.
\end{align*}
Let $T=(T_1,\dotsc,T_K)$ denote the block assignment counts, where $T_k=\sum_{i=1}^n \mathbb{I}_{\{k\}}(\hat{\tau}_i)$. 
Then the resulting prior distribution is given by
\begin{align*}
f(\tau|\theta) = \int_{\Delta_K} f(\tau|\rho)f(\rho|\theta)d\rho
 &= \frac{\Gamma(\sum_{k=1}^K \theta_k )}{\prod_{k=1}^K \Gamma(\theta_k)} \int_{\Delta_K} \left( \prod_{i=1}^n \rho_{\tau_i} \right) \left( \prod_{k=1}^K \rho_k^{\theta_k-1}\right) d\rho \\
 &= \frac{\Gamma(\sum_{k=1}^K \theta_k )}{\prod_{k=1}^K \Gamma(\theta_k)} \int_{\Delta_K} \underbrace{\prod_{k=1}^K \rho_k^{\theta_k+T_k-1}}_{\propto\mathrm{Dirichlet}(\theta+T)} d\rho \\
 &=  \frac{\Gamma(\sum_{k=1}^K \theta_k )}{\prod_{k=1}^K \Gamma(\theta_k)}  \frac{\prod_{k=1}^K \Gamma(\theta_k+T_k)}{\Gamma(n+\sum_{k=1}^K \theta_k )},
\end{align*}
which follows a Multinomial-Dirichlet distribution with parameters $\theta$ and $n$.
Therefore, the marginal posterior distribution can be expressed as
\begin{align*}
f(\tau,\nu|A)
&\propto f(A|\tau,\nu)f(\tau| \theta) f(\nu|\{\hat{\mu}_k\},\{\hat{\Sigma}_k\})\\
&\propto f(A|\tau,\nu)\left[ \prod_{k=1}^K \Gamma(\theta_k+T_k) \right]  f(\nu|\{\hat{\mu}_k\},\{\hat{\Sigma}_k\}).
\end{align*}

We can sample from the marginal posterior distribution for $\tau$ and $\nu$ via Metropolis--Hasting--within--Gibbs sampling. 
A standard Gibbs sampling update is employed to sample the posterior of $\tau$, which can be updated sequentially. The idea behind this method is to first posit a full conditional posterior distribution of $\tau$. 
Let $\tau_{-i}=\tau \setminus \tau_i$ denote the block memberships for all but vertex $i$. Then, conditioning on $\tau_{-i}$, we have
\begin{equation}\label{eqn:asgefullcond}
    f(\tau_i|\tau_{-i}, A, \nu,\theta) \propto \prod_{j\neq i}  \langle\nu_{\tau_i},\nu_{\tau_j}\rangle^{A_{ij}} (1-\langle\nu_{\tau_i},\nu_{\tau_j}\rangle)^{1-A_{ij}} \left[ \prod_{k=1}^K \Gamma(\theta_k+T_k) \right].
\end{equation}
Hence, the posterior distribution for $\tau_i \sim \mathrm{Multinomial(\rho^*_i)}$ where
\begin{equation}\label{Eqn:Gibbsasge}
  \rho^*_{i,k} = \frac{\Gamma(\theta_k+T_k) \prod_{j\neq i}  \langle\nu_{k},\nu_{\tau_j}\rangle^{A_{ij}} (1-\langle\nu_{k},\nu_{\tau_j}\rangle)^{1-A_{ij}}}
  {\sum_{k'=1}^K \Gamma(\theta_k'+T_k') \prod_{j\neq i}  \langle\nu_{k'},\nu_{\tau_j}\rangle^{A_{ij}} (1-\langle\nu_{k'},\nu_{\tau_j}\rangle)^{1-A_{ij}}}.
\end{equation}

The procedure consists of visiting each $\tau_i$ for $i = 1,\dotsc,n$ and executing Algorithm~\ref{algo:gibbs}. 
We initialize $\tau$ with $\tau^{(0)}=\hat{\tau}$, the block assignment vector obtained from GMM clustering of the estimated latent positions $\hat X$.
For the Metropolis sampler for $\nu$, the prior distribution $f(\nu|\{\hat{\mu}_k\}, \{\hat{\Sigma}_k\})$ as expressed in Eqn~(\ref{eq:ebPrior}) will be employed as the proposal distribution. 
We generate a proposed state $\tilde{\nu}_k \sim f(\nu|\{\hat{\mu}_k\}, \{\hat{\Sigma}_k\})$ with the acceptance probability defined as
\[ \min \left\{ \frac{f(A|\tau,\tilde{\nu}_k)}{f(A|\tau,\nu_k)},1\right\},\]
where $\nu_k$ in the denominator denotes the current state. 
The initialization of $\nu$ is $\nu^{(0)}| \{\hat{\mu}_k\}, \{\hat{\Sigma}_k\} \sim f(\nu| \{\hat{\mu}_k\}, \{\hat{\Sigma}_k\})$.

\begin{algorithm}[H]
\caption{Metropolis--Hasting--within--Gibbs sampling\\for the block membership vector $\tau$ and the latent positions $\nu_1,\cdots,\nu_K$}
\label{algo:gibbs}
\begin{algorithmic}[1]
        \STATE At iteration $h$;
        \FOR {$i=1$ to $n$}
        \STATE Compute $\rho^*_i(\tau^{(h)}_1,\dotsc,\tau^{(h)}_{i-1},\tau^{(h-1)}_{i+1},\tau^{(h-1)}_{n})$ as in Eqn~(\ref{Eqn:Gibbsasge})
        \STATE Set $\tau^{(h)}_i = k$ with probability $\rho^*_{i,k}$ 
        \ENDFOR
        \STATE Generate $\tilde{\nu} \sim \mathbb{I}_{\mathcal{S}}(\nu) \prod_{k=1}^K \mathcal{N}_d(\nu_k \mid \hat{\mu}_k, \hat{\Sigma}_k)$
        \STATE Compute the acceptance probability $\pi(\tilde{\nu}) = \mathrm{min}\{1,\frac{f(A|\tau^{(h)},\tilde{\nu})}{f(A|\tau^{(h)},\nu^{(h-1)})}\}$
        \STATE Set \[ \nu^{(h)} = \begin{cases}
        \tilde{\nu} & \text{with probability}\;  \pi(\tilde{\nu})\\
        \nu^{(h-1)} & \text{with probability}\;  1-\pi(\tilde{\nu})
        \end{cases}\]
\end{algorithmic}
\end{algorithm}

\subsection{The Alternative ``Flat'' Model}\label{subsubsection:flat}
In the event that no special prior information is available, a natural choice of prior is the uniform distribution on the parameter space. This results in the formulation of the \emph{Flat} model as an alternative to an empirical Bayes prior distribution for $\nu$ discussed in the previous section. 
We consider a flat prior distribution on the constraint set $\mathcal{S}$, 
where the marginal posterior distribution for $\tau$ and $\nu$ is given by
\begin{align*}
f(\tau,\nu|A)
&\propto f(A|\tau,\nu)f(\tau| \theta) f(\nu)\\
&\propto f(A|\tau,\nu)\left[ \prod_{k=1}^K \Gamma(\theta_k+T_k) \right]  \mathbb{I}_\mathcal{S}(\nu).
\end{align*}
The Gibbs sampler for $\tau$ is identical to the procedure presented in Section~\ref{ebsbm}. As for the Metropolis sampler for the latent positions $\nu$, the flat prior distribution is used as the proposal. However, we initialize $\nu$ by generating it from the prior distribution of $\nu$ as \emph{ASGE}, i.e. $f(\nu|\{\hat{\mu}_k\}, \{\hat{\Sigma}_k\})$. 

\subsection{Comparison Benchmarks}\label{alternatives}
\subsubsection*{``Exact''}\label{subsubsection:exact}
This is our primary benchmark where the latent positions $\nu$ and the block membership probabilities $\rho$ are assumed known. Thus, the posterior distribution for the block memberships $\tau$ is given by 
\begin{align*}
f(\tau|\: A,\nu,\rho) &\propto f(A|\tau,\nu) f(\tau|\rho)\\
&= \prod_{i=1}^n \rho_{\tau_{i}} \prod_{i<j} \langle\nu_{\tau_i},\nu_{\tau_j}\rangle^{A_{ij}} (1-\langle\nu_{\tau_i},\nu_{\tau_j}\rangle)^{1-A_{ij}} .
\end{align*}
We can draw inferences about $\tau$ based on the posterior $f(\tau| A,\nu,\rho)$ via an \emph{Exact} Gibbs sampler using its full-conditional distribution,
\begin{equation}
  f(\tau_i|\tau_{-i}, A, \nu,\rho) \propto \rho_{\tau_i} \prod_{j\neq i}  \langle\nu_{\tau_i},\nu_{\tau_j}\rangle^{A_{ij}} (1-\langle\nu_{\tau_i},\nu_{\tau_j}\rangle)^{1-A_{ij}},
\end{equation}
which is the multinomial$(\rho^*_i)$ density where
\begin{equation}\label{Eqn:Gibbsprob}
  \rho^*_{i,k} = \frac{\rho_{k} \prod_{j\neq i}  \langle\nu_{k},\nu_{\tau_j}\rangle^{A_{ij}} (1-\langle\nu_{k},\nu_{\tau_j}\rangle)^{1-A_{ij}}}
  {\sum_{k'=1}^K \rho_{k'} \prod_{j\neq i}  \langle\nu_{k'},\nu_{\tau_j}\rangle^{A_{ij}} (1-\langle\nu_{k'},\nu_{\tau_j}\rangle)^{1-A_{ij}}}.
\end{equation}
Hence, for our \emph{Exact} Gibbs sampler, once a vertex is selected, the exact calculation of $\rho^{(i)}$ and sample $\tau_i$ from the $\mathrm{Multinomial}(\rho^{(i)})$ can easily be obtained. 
Initialization of $\tau$ will be $\tau^{(0)}_1,\dotsc, \tau^{(0)}_n|\rho \stackrel{iid}{\sim} \mathrm{Multinomial}(\rho)$.

\subsubsection*{``Gold''}\label{subsubsection:gold}
For our secondary benchmark, we assume $\rho$ is known and that both $\nu$ and $\tau$ are unknown. 
Here we describe what we call the {\em Gold} standard prior distribution. 

Let the true value for the latent positions be represented by $\nu^{*}$.
Based on Corollary~\ref{cor:sbmCLT}, we can suppose that the prior distribution for $\nu_k$ follows a (truncated) multivariate Gaussian centered at $\nu_k^*$ and with covariance matrix $\Sigma^*_k = (1/n)\Sigma_k$ given by the theoretical limiting distribution for the adjacency spectral embedding $\hat{X}$ presented in Eqn~(\ref{eq:condConv}) (i.e. $\nu|\{\nu^*_k\},\{\Sigma^*_k\} \sim \mathcal{N}_d\left(\nu_k|\nu^*_k,\Sigma^*_k \right) $). This corresponds to the approximate distribution of $\hat{X}_i$ if we condition on $\tau_i=k$. This gold standard prior can be thought of as an oracle; however, in practice the theoretical $\nu^{*}$ and $\Sigma^*_k$ are not available.

Inference for $\tau$ and $\nu$ is based on the posterior distribution, $f(\tau,\nu|A,\rho)$, estimated by samples obtained from a Gibbs sampler for $\tau$ and an Independent Metropolis sampler for $\nu$. Thus, the posterior distribution for the unknown quantities is given by
\begin{align*}
f(\tau,\nu|A,\rho) &\propto f(A|\tau,\nu)f(\tau|\rho) f(\nu|\{\nu^*_k\},\{\Sigma^*_k\})\\
&= \left[ \prod_{i=1}^n \rho_{\tau_{i}} \prod_{i<j} \langle\nu_{\tau_i},\nu_{\tau_j}\rangle^{A_{ij}} (1-\langle\nu_{\tau_i},\nu_{\tau_j}\rangle)^{1-A_{ij}}   \right] f(\nu|\{\nu^*_k\}, \{\Sigma^*_k\}),
\end{align*}
In this case, the Gibbs sampler for $\tau$ will be identical to that for \emph{Exact} except the initial state $\tau^{(0)}$ will be given by $\hat{\tau}$, the block assignment vector obtained from GMM as explained in Section~\ref{ebsbm}. Similar to the \textit{ASGE} model, the prior distribution $f(\nu|\{\nu^*_k\},\{\Sigma^*_k\})$ will be employed as the proposal distribution for the Metropolis sampler for $\nu$.

Table~\ref{tab:samples} provides a summary of our Bayesian modeling schemes. The adjacency spectral graph embedding theory suggests that we might expect increasingly better performance as we go from \emph{Flat} to \emph{ASGE} to \emph{Gold} to \emph{Exact}. (As a teaser, we hint here that we will indeed see precisely this progression, in Section~\ref{section:eg}.)

\begin{table}[H]
\caption{Bayesian Sampling Schemes}
\label{tab:samples}
\begin{center}
\begin{tabular}{p{0.1\textwidth}clllll}
\addlinespace
\toprule
\multicolumn{3}{c}{\textbf{Models}} & \textbf{Exact}& \textbf{Gold}& \textbf{ASGE} & \textbf{Flat}\\
\cmidrule(r){1-3}
&{\textbf{Parameters}}& \\
\midrule
\parbox[t]{2mm}{\multirow{2}{*}{\textbf{Gibbs}}}
& $\bm{\tau_i}$
&  Prior
& $\tau_{i}|\rho \sim$
& $\tau_i|\rho \sim $
& $T|\theta \sim $
& $T|\theta \sim $ \\
&&& $\mathrm{Multinomial}(\rho)$
& $\mathrm{Multinomial}(\rho)$
& $\mathrm{Multinomial-}$
& $\mathrm{Multinomial-}$\\
&&&&& $\mathrm{Dirichlet}(\theta,n)$
& $\mathrm{Dirichlet}(\theta,n)$\\
\addlinespace
\addlinespace
&&  Initial Point
& $\tau_i|\rho \sim $ & $\hat{\tau}$ & $\hat{\tau}$ & $\hat{\tau}$ \\
&&& $\mathrm{Multinomial}(\rho)$ &&&\\
\addlinespace
\midrule
\multirow{3}{\linewidth}{\textbf{Independent Metropolis Hasting}}& $\bm{\nu_k}$ & Prior & $-$    & $\nu_k | \nu^*_k,\Sigma^*_k \sim $   &   $\nu_k | \hat{\mu}_k,\hat{\Sigma}_k  \sim $ & $\nu_k \sim$ \\
& & & & $\mathcal{N}(\nu^*_k,\Sigma^*_k)$ & $\mathcal{N}(\hat{\mu}_k,\hat{\Sigma}_k)$ & $ \mathcal{U}(\mathcal{S})$\\
\addlinespace
\addlinespace
 &  & Initial point & $-$   & $\nu^{(0)}_k| \nu^*_k,\Sigma^*_k \sim $  & $\nu^{(0)}_k | \hat{\mu}_k,\hat{\Sigma}_k \sim $ & $\nu^{(0)}_k |\sim $\\
&&&& $\mathcal{N}(\nu^*_k,\Sigma^*_k)$ & $\mathcal{N}(\hat{\mu}_k,\hat{\Sigma}_k)$ & $\mathcal{N}(\hat{\mu}_k,\hat{\Sigma}_k)$\\
\addlinespace
\addlinespace
& & Proposal & $-$& $\tilde{\nu}_k| \nu^*_k,\Sigma^*_k \sim $ & $\tilde{\nu}_k| \hat{\mu}_k,\hat{\Sigma}_k \sim $ & $\tilde{\nu}_k \sim $ \\
&&&& $\mathcal{N}(\nu^*_k,\Sigma^*_k)$ & $\mathcal{N}(\hat{\mu}_k,\hat{\Sigma}_k)$ & $\mathcal{U}(\mathcal{S})$ \\
\bottomrule
\end{tabular}
\end{center}
\end{table}

\section{Performance Comparisons}\label{section:eg}

We illustrate the performance of our \emph{ASGE} model via various Monte Carlo simulation experiments and one real data experiment.
Specifically, we consider 
in Section~\ref{subsec:2blocks} a $K=2$ SBM,
in Section~\ref{subsec:2blocksD} a generalization of this $K=2$ SBM to a more general RDPG, 
in Section~\ref{subsec:3blocks} a $K=3$ SBM,
and in Section~\ref{subsec:wiki} a three-class Wikipedia graph example.
We demonstrate the utility of the \emph{ASGE} model for estimating vertex block assignments via comparison to competing methods.

Throughout our performance analysis,
we generate posterior samples of $\tau$ and $\nu$ for a large number of iterations for two parallel Markov chains. 
The percentage of misassigned vertices per iteration is calculated and used to compute Gelman-Rubin statistics to check convergence of the chains. 
The posterior inference for $\tau$ is based on iterations after convergence.
Performance is evaluated by calculating the vertex block assignment error. 
This procedure is repeated multiple times to obtain estimates of the error rates. 

\subsection{A Simulation Example with $K = 2$}\label{subsec:2blocks}

Consider the SBM parameterized by

\begin{equation}
\label{eq:SBM}
{B} = \left(
      \begin{array}{cc}
        0.42 & 0.42 \\
        0.42 & 0.5 \\
      \end{array}
    \right) \qquad\mathrm{and}\qquad  \rho = (0.6,0.4).
\end{equation}

The block proportion vector $\rho$ indicates that each vertex will be in block 1 with probability $\rho_1=0.6$ and in block 2 with probability $\rho_2=0.4$. 
Edge probabilities are determined by the entries of $B$, independent and a function of only the vertex block memberships.
This model can be parameterized as an RDPG in $\Re^2$ 
where the distribution $F$ of the latent positions 
is a mixture of point masses positioned at 
$\nu_1 \approx (0.5489,0.3446)$ with prior probability 0.6 and $\nu_2 \approx (0.3984,0.5842)$ with prior probability 0.4. 

For each $n \in \{100,250,500,750,1000\}$, 
we generate random graphs according to the SBM with parameters as provided in Eqn~(\ref{eq:SBM}). 
For each graph $G$, the spectral decomposition of the corresponding adjacency matrix $A$ 
provides the estimated latent positions $\hX$. 

Subsequently, GMM is used to cluster the embedded vertices, the result of which (estimated block memberships $\hat{\tau}$ derived from the individual mixture component membership probabilities from the estimated Gaussian mixture model) is then reported as GMM performance as well as employed as the initial point in the Gibbs step for updating $\tau$. The mixture component means $\hat{\mu}_k$ and variances $\hat{\Sigma}_k$ determine our empirical Bayes \emph{ASGE} prior for the latent positions $\nu$. 
The GMM estimate of block proportion vector $\hat \rho$ in place of a conjugate Dirichlet prior on $\rho$ was considered, but no substantial performance improvements were realized.
To avoid the model selection quagmire we assume $d=2$ and $K=2$ are known in this experiment. 
\begin{figure}[htpb]
\begin{subfigure}[b]{0.5\linewidth}
\centering
\includegraphics[width=\textwidth]{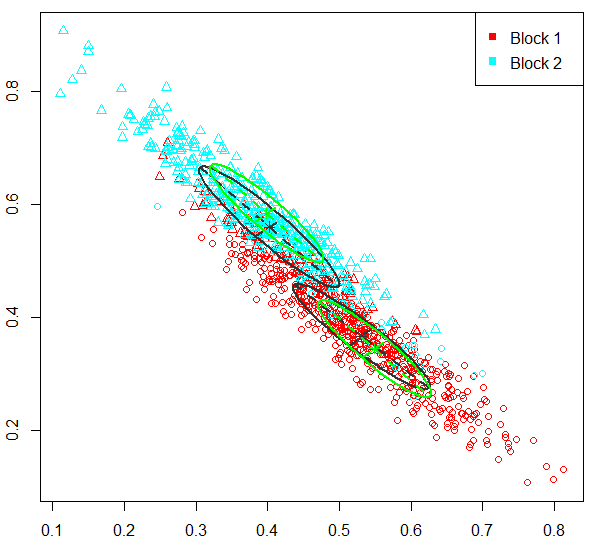}
\end{subfigure}
\begin{subfigure}[b]{0.5\linewidth}
\centering
\includegraphics[width=\textwidth]{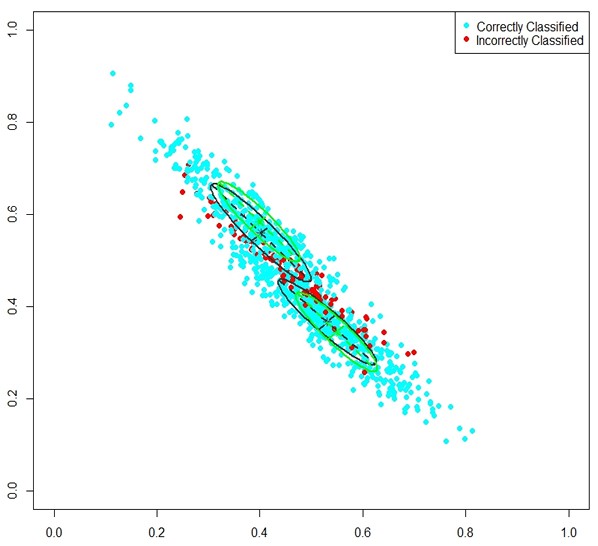}
\end{subfigure}
\caption{\label{fig:mclust_n1000}Scatter plot of the estimated latent positions 
 $\hX_i$ for one Monte Carlo replicate with $n=1000$ for the $K=2$ SBM considered in Section~\ref{subsec:2blocks}.
In the left panel, the colors denote the true block memberships for the corresponding vertices in the SBM, while the symbols denote the cluster memberships given by the GMM. In the right panel, the colors represents whether the vertices are correctly or incorrectly classified by the \textit{ASGE} model. The ellipses represent the 95\% level curves of the estimated GMM (black) and the theoretical GMM (green). Note that misclassification occurs where the clusters are overlapping. 
}
\end{figure}

Figure~\ref{fig:mclust_n1000} presents a scatter plot of the estimated latent positions $\hX_i$ for one Monte Carlo replicate with $n=1000$.
The colors denote the true block memberships for the corresponding vertices in the SBM. 
The symbols denote the cluster memberships given by the GMM.
The ellipses represent the 95\% level curves of the estimated GMM (black) and the theoretical GMM (green).
\begin{figure}[htpb]
\centering
\includegraphics[width=0.7\linewidth]{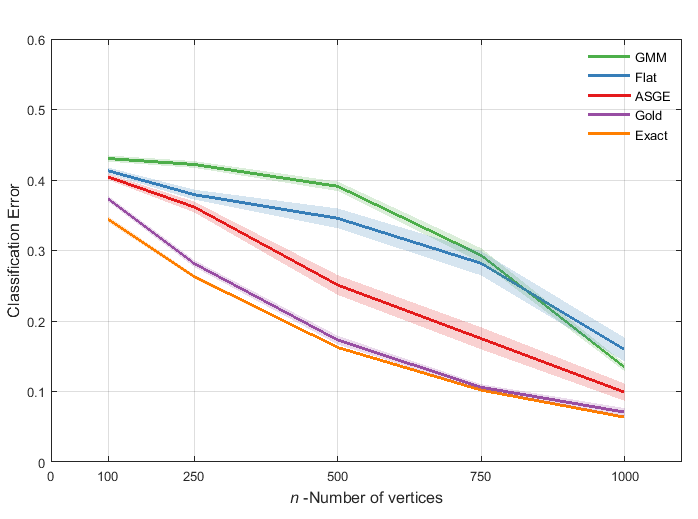}
\caption{Comparison of vertex block assignment methodologies for the $K=2$ SBM considered in Section~\ref{subsec:2blocks}.
Shaded areas represent standard errors.
The plot indicates that utilizing a multivariate Gaussian mixture estimate for the estimated latent positions
as an empirical Bayes prior (\emph{ASGE}) can yield substantial improvement over both the GMM vertex assignment and the Bayesian method with a \emph{Flat} prior.
See text for details and analysis.}
\label{fig:misclass}
\end{figure}
Results comparing with the alternative \textit{Flat}, benchmark models, and GMM are presented in Figure~\ref{fig:misclass}. As expected, the error decreases for all models as the number of vertices $n$ increases. As previously explained in Section~\ref{models}, \emph{Exact} and \emph{Gold} formulated in this study are perceived as benchmarks;
it is expected that these models will show the best performance
--
for \emph{Exact}, all the parameters are assumed known apart from the block memberships $\tau$, 
while in the case of the \emph{Gold} model, although the latent positions $\nu$ and $\tau$ are unknown parameters, 
their prior distributions were taken from the true latent positions and the theoretical limiting covariances. 

The main message from Figure~\ref{fig:misclass}
is that our empirical Bayes model, \emph{ASGE}, is vastly superior to that of both the alternative \emph{Flat} model and GMM (the sign test $p$-value for the paired Monte Carlo replicates is less than $10^{-10}$ for both comparisons for all $n$)
and nearly achieves \emph{Gold}/\emph{Exact} performance by $n=1000$.
As an aside, we note that when we put a flat prior directly on $B$, 
we obtain results indistinguishable from our \emph{Flat} model on the latent positions. 

\begin{figure}[htpb]
\centering
\includegraphics[width=0.7\linewidth]{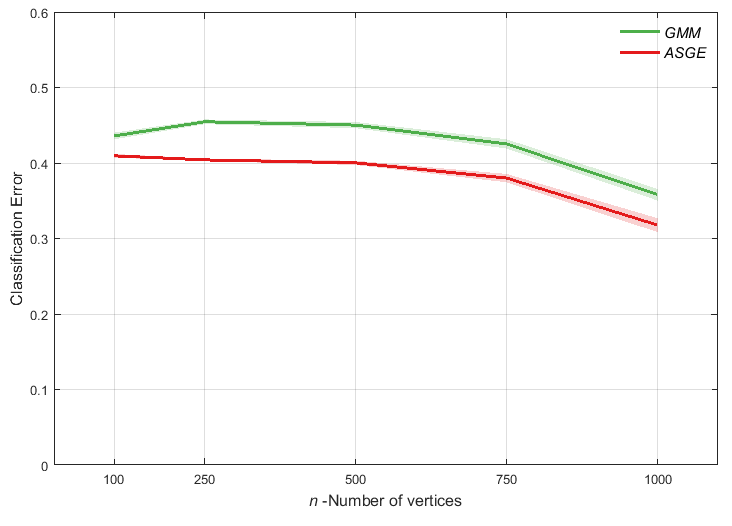}
\caption{Comparison of classification error for GMM and \textit{ASGE} in the sparse simulation setting. Shaded areas denote standard errors. The plot suggests that we obtain similar comparative results, with analogous ASGE superiority, in a sparse simulation setting.
}
\label{fig:sparsesim}
\end{figure}

A version of Theorem~\ref{thm:clt} for sparse random dot product graphs is given in~\citet{sussman2014foundations}, and suggests an empirical Bayes prior for use in sparse graphs. A thorough investigation of comparative performance in this case is beyond the scope of this manuscript, but we have provided illustrative results in Figure~\ref{fig:sparsesim} for the sparse graphs analogous to the setting presented in Eqn~(\ref{eq:SBM}). For the same values of $n$, we generate sparse random graphs from the following SBM:
\begin{equation*}
\label{eq:SBMsparse}
{B} = \left(
      \begin{array}{cc}
        0.42 & 0.2 \\
        0.2 & 0.5 \\
      \end{array}
    \right)*\frac{1}{\sqrt{n}}  \qquad\mathrm{and}\qquad  \rho = (0.6,0.4).
\end{equation*}
For clarity, the plot includes only $ASGE$ and GMM. Note that similar performance gains are obtained, with analogous \textit{ASGE} superiority, in the sparse simulation setting (although absolute performance is of course degraded).

\subsection{A Dirichlet Mixture RDPG Generalization}\label{subsec:2blocksD}

Here we generalize the simulation setting presented in Section~\ref{subsec:2blocks} above
to the case where the latent position vectors
are distributed according to a mixture of Dirichlets
as opposed to the SBM's mixture of point masses.
That is, we consider $X_i \stackrel{iid}{\sim} \sum_k \rho_k Dirichlet(r \cdot \nu_k)$.
Note that the SBM model presented in Section~\ref{subsec:2blocks} is equivalent to the limit of this mixture of Dirichlets model as $r \to \infty$.

For $n=500$, we report illustrative results using $r=100$,
for comparison with the SBM results from Section~\ref{subsec:2blocks}.
Specifically, we obtain mean error rates of
0.4194, 0.2865, and 0.3705
for
\emph{Flat}, \emph{ASGE}, and GMM, respectively;
the corresponding results for the SBM, from Figure~\ref{fig:misclass}, are
0.3456, 0.2510, and 0.3910.
Thus we see that, while the performance is slightly degraded,
 our empirical Bayes approach works well in this RDPG generalization of Section~\ref{subsec:2blocks}'s SBM.
This demonstrates robustness of our method to violation of the SBM assumption.

\subsection{A Simulation Example with $K = 3$}\label{subsec:3blocks}

Our final simulation study considers the $K=3$ SBM
parameterized by

\begin{equation}
\label{eq:3SBM}
B = \left(
      \begin{array}{ccc}
        0.6 & 0.4 & 0.4 \\
        0.4 & 0.6 & 0.4 \\
        0.4 & 0.4 & 0.6 \\
      \end{array}
    \right) \qquad\mathrm{and}\qquad  \rho = (1/3,1/3,1/3).
\end{equation}

In a same manner as 
Section~\ref{subsec:2blocks}, 
the model is parameterized as an RDPG in $\mathbb{R}^3$ where the distribution of the latent positions is a mixture of point masses positioned at $\nu_1 \approx (0.68,0.20,-0.30)$, $\nu_2 \approx (0.68,-0.36,-0.02)$, $\nu_3 \approx (0.68,0.16,0.33)$ with equal probabilities. In this experiment, we assume that $d=3$ and $K=3$ are known.

Table~\ref{tab:error_rates_K3} displays error rate estimates for this case, with $n=150$ and $n=300$. 
In both cases, the \emph{ASGE} model yields results vastly superior to the \emph{Flat} model;
e.g., for $n=300$ the mean error rate for \emph{Flat} is approximately 11\%
compared to a mean error rate for \emph{ASGE} of approximately 1\%.
Based on the paired samples, 
the sign test $p$-value is less than $10^{-10}$ for both values of $n$. 
While the results of GMM appear competitive to the results of our empirical Bayes with \emph{ASGE} prior in terms of mean and median error rate, the paired analysis shows again that the \emph{ASGE} prior is superior, as seen by sign test $p$-values < $10^{-10}$ for both values of $n$.

\begin{figure}[H]
\centering
\includegraphics[scale=1.0]{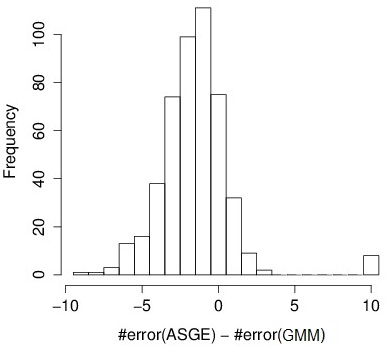}
\caption{Histogram (500 Monte Carlo replicates) 
of the differential number of errors made by \emph{ASGE} and GMM 
 for the $K=3$ SBM considered in Section~\ref{subsec:3blocks}, with $n=300$,
indicating the superiority of \emph{ASGE} over GMM.
For most graphs, emprical Bayes with \emph{ASGE} prior performs as well as or better than GMM --
the sign test for this paired sample yields $p \approx 0$.}
\label{fig:histn300}
\end{figure}

From Table~\ref{tab:error_rates_K3}, we see that for $n=300$, empirical Bayes with \emph{ASGE} prior has a mean error rate of $1$ percent (3 errors per graph) and a median error rate of 1/3 percent (1 error per graph), while GMM has a mean and median error rate of 1 percent. As an illustration, Figure~\ref{fig:histn300} presents the histogram of the differential number of errors made by the \emph{ASGE} model and GMM for $n=300$. 
The histogram shows that for most graphs, empirical Bayes with \emph{ASGE} prior performs as well as or better than GMM.
(NB: In the histogram presented in Figure~\ref{fig:histn300}, 
eight ouliers in which \emph{ASGE} performed particularly poorly are censored at a value of 10;
we believe these outliers are due to chain convergence issues.

\begin{table}[H]
\centering
\begin{tabular}{lp{6em}ccc}
\toprule
\addlinespace
 & & $n=150$ & $n=300$\\
 \midrule
 \addlinespace
$\widehat{L}$(Flat) & mean & 0.3288 & 0.1137 \\
 & 95\% CI & [0.3207,0.3369] & [0.1004,0.1270] \\
 & median & 0.3600 & 0.0133 \\
 \midrule
  \addlinespace
$\widehat{L}$(ASGE) & mean & 0.1359 & 0.0107 \\
 & 95\% CI & [0.1277,0.1440] & [0.0069,0.0145] \\
 & median & 0.0733 & 0.0033 \\
 \midrule
  \addlinespace
$\widehat{L}$(GMM) & mean & 0.1438 & 0.0110 \\
 & 95\% CI & [0.1396,0.1480] & [0.0104,0.0116] \\
 & median & 0.1267 & 0.0100 \\
 \bottomrule
\end{tabular}
\caption{\label{tab:error_rates_K3}Error rate estimates for the $K=3$ SBM considered in Section~\ref{subsec:3blocks}.}
\end{table}

\subsection{Wiki Experiment} \label{subsec:wiki}

In this section we analyze an application of our methodology to the Wikipedia graph.
The vertices of this graph represent Wikipedia article pages
 and there is an edge between two vertices if either of the associated pages hyperlinks to the other.
The full data set consists of 1382 vertices -- the induced subgraph generated from the two-hop neighborhood of the page ``Algebraic Geometry.''
Each vertex is categorized by hand into one of six classes
  -- $People$, $Places$, $Dates$, $Things$, $Math$, and $Categories$ --
based on the content of the associated article.
(The adjacency matrix and the true class labels for this data set are available at \url{http://www.cis.jhu.edu/~parky/Data/data.html}.)

\begin{figure}[H]
\centering
\includegraphics[width=0.6\textwidth]{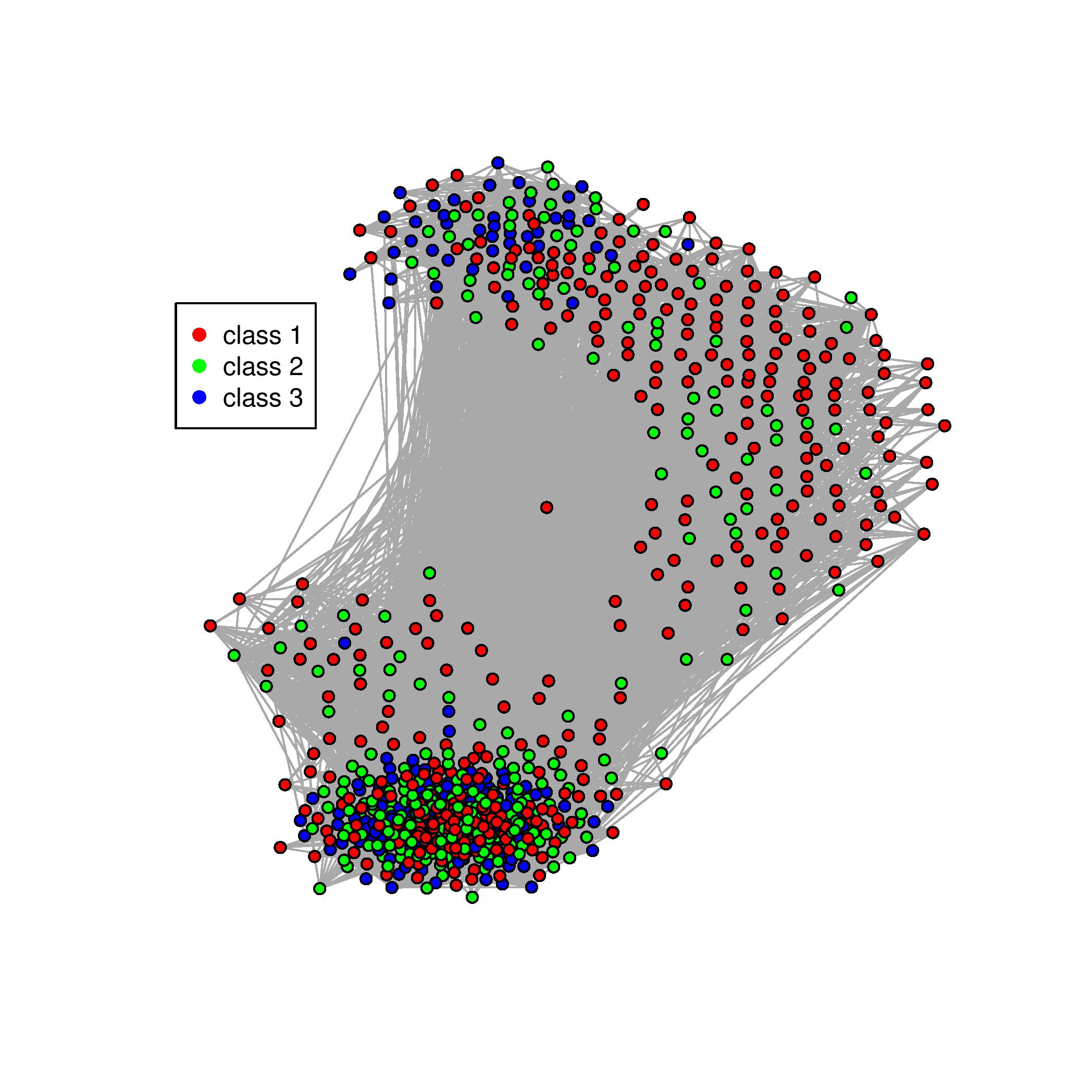}
\caption{\label{fig:graph}
Our Wikipedia graph, with $m=828$ vertices:
$m_1 = 368$ for Class 1 = $People$ = red;
$m_2 = 269$ for Class 2 = $Places$ = green;
$m_3 = 191$ for Class 3 = $Dates$ = blue.}
\end{figure}

\begin{figure}[H]
\centering
\includegraphics[width=0.6\textwidth]{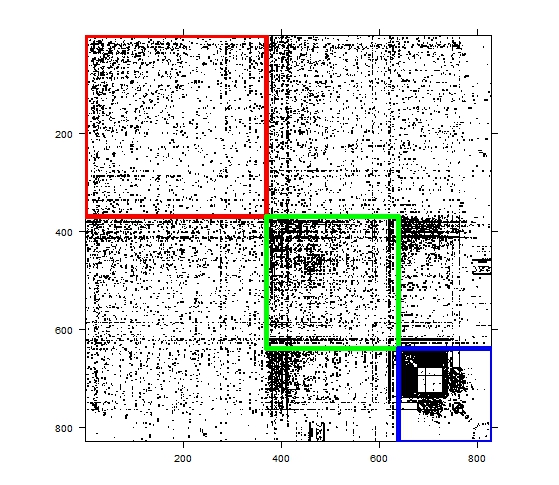}
\caption{\label{fig:imageA}
The adjacency matrix for our Wikipedia graph.}
\end{figure}

\begin{figure}[H]
\centering
\includegraphics[width=0.6\textwidth]{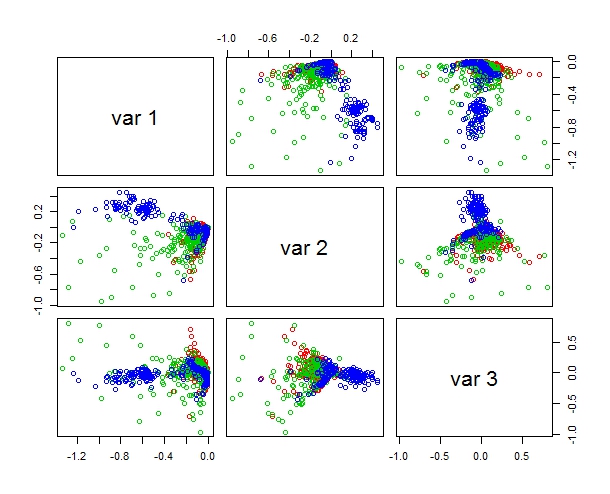}
\caption{\label{fig:Origin1v2v3}
The adjacency spectral embedding for our Wikipedia graph.}
\end{figure}

We analyze a subset of this data set corresponding to the $K=3$ classes $People$, $Places$, and $Dates$,
 labeled here as Class 1, 2 and 3, respectively.
After excluding three isolated vertices in the induced subgraph generated by these three classes,
we have a connected graph with a total of $m = 828$ vertices;
the class-conditional sample sizes are $m_1 = 368$, $m_2 = 269$, and $m_3 = 191$.
Figure~\ref{fig:graph} presents one rendering of this graph
(obtained via one of the standard force-directed graph layout methods, using the command \texttt{layout.drl} in the \texttt{igraph} R package);
Figure~\ref{fig:imageA} presents the adjacency matrix;   
Figure~\ref{fig:Origin1v2v3} presents the pairs plot for the adjacency spectral embedding of this graph into $\mathbb{R}^3$.
(In all figures, we use red for Class 1, green for Class 2 and blue for Class 3.)
Figures 
\ref{fig:graph},
\ref{fig:imageA}, and
\ref{fig:Origin1v2v3} 
indicate clearly that this Wikipedia graph is {\em not} a pristine SBM -- real data will never be;
nonetheless, we proceed undaunted.  

We illustrate our empirical Bayes methodology, following Algorithm~\ref{algo:EB}, via a bootstrap experiment. We generate bootstrap resamples from the adjacency spectral embedding $\hat X$
depicted in Figure~\ref{fig:Origin1v2v3},
with $n=300$ ($n_1 = n_2 = n_3 = 100$). 
This yields $\widehat{X}^{(b)}$ for each bootstrap resample $b = 1,\dots,200$.
It is important to note that we regenerate an RDPG based on the sampled latent positions, and proceed from this graph with our full empirical Bayes analysis, for each resample. This provides for valid inference conditional on the $\hat{X}$ --
that is, this bootstrap procedure is justified for confidence intervals assuming the true latent positions are $\hat{X}$, and provides for unconditional inference only asymptotically as $\hat{X} \to X$.

\begin{figure}[H]
\centering
\includegraphics[width=0.6\textwidth]{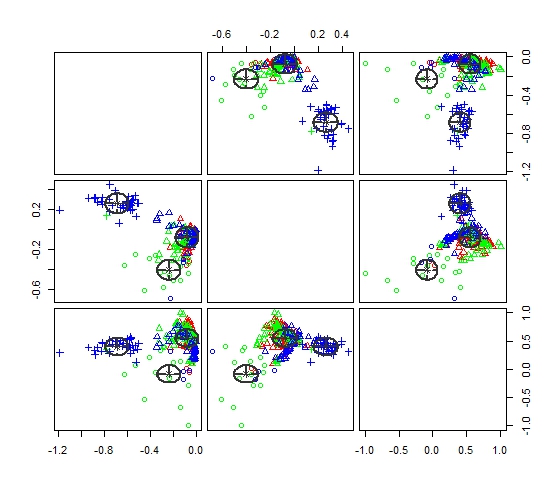}
\caption{\label{fig:MclustFancy}
Illustrative empirical prior for one bootstrap resample ($n=300$) for our Wikipedia experiment;
colors represent true classes, $K=3$ estimated Gaussians are depicted with level curves, and symbols represent GMM cluster memberships.}
\end{figure}

As before, GMM
is used to cluster the (embedded) vertices 
and obtain block label estimates $\hat{\tau}$ 
and mixture component means $\widehat{\mu}_k$ and variances $\widehat{\Sigma}_k$ 
for each cluster $k$ of the estimated latent positions $\widehat{X}^{(b)}$.
The clustering result from GMM for one resample is presented in Figure~\ref{fig:MclustFancy}.
(We choose $d=3$ for the adjacency spectral embedding dimension
because a common and reasonable choice is to use $d=K$,
which choice is justified in the SBM case \citep{fishkind2013consistent}.)
The GMM clustering
provides the empirical prior and starting point for our Metropolis--Hasting--within--Gibbs sampling
(Algorithm~\ref{algo:gibbs})
using the subgraph of the full Wikipedia graph induced by $\hat{X}^{(b)}$.  
(NB: For this Wikipedia experiment, the assumption of homophily is clearly violated;
as a result, the constraint set used here is given by
$\mathcal{S} = \{ \nu \in \mathcal{R}^{K\times d}: \forall i,j \in [K], 0 \le \left \langle \nu_i,\nu_j \right \rangle \le 1 \}$.)

\begin{figure}[H]
\centering
\includegraphics[width=0.5\textwidth]{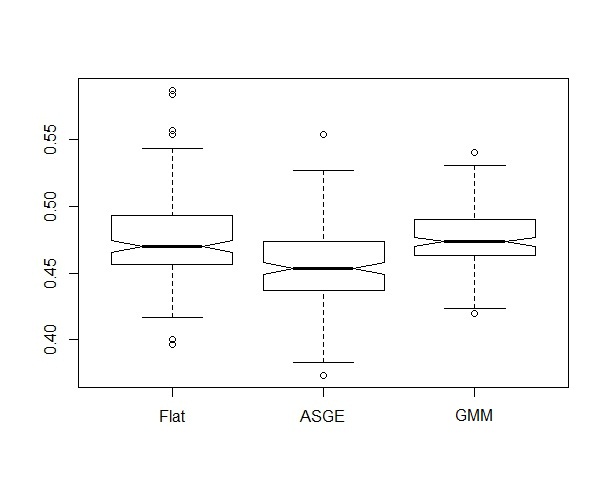}
\caption{\label{fig:boxplot}
Boxplot of classification errors for our Wikipedia experiment.}
\end{figure}

Classification results for this experiment are depicted via boxplots in Figure~\ref{fig:boxplot}.
We see from the boxplots that using the adjacency spectral empirical prior does yield statistically significant improvement;
indeed, our paired sample analysis yields sign test $p$-values less than $10^{-10}$ for both \emph{ASGE} vs \emph{Flat} and \emph{ASGE} vs GMM. Notably, \textit{ASGE} and \textit{Flat} differ by $9.35\%$ in average, which is approximately 28 different classifications per graph. Despite similar predictions, \textit{ASGE} improves \textit{Flat}.

We have shown that using the empirical \emph{ASGE} prior has improved performance compared to the \emph{Flat} prior and GMM on this Wikipedia dataset.
However, Figure~\ref{fig:boxplot} also indicates that \emph{ASGE} performance on this data set,
while representing a statistically significant improvement,
 might seem not very good in absolute terms:
the mean probability of misclassification over bootstrap resamples is
 $\widehat{L} \approx 0.456$ for \emph{ASGE} versus
 $\widehat{L} \approx 0.476$ for both \emph{Flat} and GMM.
That is, empirical Bayes using the adjacency spectral prior provides a 
statistically significant but perhaps unimpressive 2\% improvement in the error rate.
(Note that chance performance is $L = 2/3$.)
Given that the Bayes optimal probability of misclassification $L^*$ is unknown,
we consider $\inf_{\phi \in \mathcal{C}} L(\phi)$
where $\mathcal{C}$ denotes the class of all classifiers based on class-conditional Gaussians.
This yields an error rate of approximately $0.401$.
Note that this analysis assumes that a training set of $n=300$ labeled exemplars is available,
which training information is {\em not} available in our empirical Bayes setting.
Nonetheless, we see that our empirical Bayes methodology using the \emph{ASGE} prior
improves more than 25\% of the way from the \emph{Flat} and GMM performance to this (presumably unattainable) standard.
As a final point, we note that a $k$-nearest neighbor classifier (again, with a training set of $n=300$ labeled exemplars)
yields an error rate of approximately $0.338$,
indicating that the assumption of class-conditional Gaussians was unwarranted.
(Indeed, this is clear from Figure~\ref{fig:Origin1v2v3}.)
That our \emph{ASGE} provides significant performance improvement 
despite the fact that our real Wikipedia data set so dramatically violates the stochastic block model assumptions
is a convincing demonstration of the robustness of the methodology.

\section{Conclusion}\label{section:discussion}

In this paper we have formulated an empirical Bayes estimation approach for block membership assignment. Our methodology is motivated by recent theoretical advances regarding the distribution of 
the adjaceny spectral embedding of random dot product and SBM graphs. To apply our model, we derived a Metropolis-within-Gibbs algorithm for block membership and latent position posterior inference. 

Our simulation experiments demonstrate that the \emph{ASGE} model consistently outperforms the GMM clustering used as our emprical prior as well as the alternative \emph{Flat} prior model --
notably, even in our Dirichlet mixture RDPG model wherein the SBM assumption is violated. 
For the Wikipedia graph, our \emph{ASGE} model again performs admirably,
even though this real data set is far from an SBM. 
Our results focus on demonstrating the utility of the~\citet{athreya2013limit} limit theorem for an SBM in providing an empirical Bayes prior as a mixture of Gaussians. Although there are myriad non-adjacency spectral embedding approaches, for ease of comparison we instead consider different Bayes samplers. One promising comparison for future investigation involves profile likelihood methods,
which can potentially produce estimates akin to our maximum likelihood mixture estimates.

We considered only simple graphs; extension to directed and weighted graphs is of both theoretical and practical interest.

To avoid the model selection quagmire, we have assumed throughout that the number of blocks $K$ and the dimension of the latent positions $d$ are known.
Model selection is in general a difficult problem;
however, automatic determination of both the dimension $d$ for a truncated eigen-decomposition and the complexity $K$ for a Gaussian mixture model estimate
are important practical problems and thus have received enormous attention in both the theoretical and applied literature.
For our case, \citet{fishkind2013consistent} demonstrates that the SBM embedding dimension $d$ can be successfully estimated, and \citet{FRmclust} provides one common approach to estimating the number of Gaussian mixture components $K$.
We note that $d=K$ is justified for the adjacency spectral embedding dimension of an SBM, as increasing $d$ beyond the true latent position dimension adds variance without a concomitant reduction in bias. It may be productive to investigate {\it simultaneous} model selection methodologies for $d$ and $K$. Moreover, robustness of the empirical Bayes methodology to misspecification of $d$ and $K$ is also of great practical importance. 

In the dense regime, raw spectral embedding even without the empirical Bayes augmentation does provide strongly consistent classification and clustering~\citep{lyzinski2013perfect,sussman2012consistent}. However, this does not rule out the possibility of substantial performance gains for finite sample sizes. It is these finite sample performance gains that are the main topic of this work, and that we have demonstrated conclusively. We note that while~\citet{sussman2014foundations} provides a non-dense version of the CLT, briefly discussed in this paper, both theoretical and methodological issues remain in developing its utility for generating an empirical prior. This is of considerable interest and thus a more comprehensive understanding of the CLT for non-dense RDPGs is a priority for ongoing research. 

Additionally, we computed Gelman-Rubin statistics based on the percentage of misclassified vertices per iteration to check convergence of the MCMC chains. For large number of vertices $n$, where perfect classification is obtainable, this diagnostic will fail; however for cases of interest (in general, and specifically in this work) in which perfect classification is beyond reasonable expectation and empirical Bayes improves performance, this diagnostic is viable.

Finally, we note that we have made heavy use of the dot product kernel.
\citet{tang2013} provides some useful results for the case of a latent position model with unknown kernel, but we see extending our empirical Bayes methodology to this case as a formidable challenge. Recent results on the SBM as a universal approximation to general latent position graphs \citep{NIPS2013_5047,2013arXiv1312.5306O}
suggest, however, that this challenge, once surmounted, may provide a simple consistent framework for empirical Bayes inference on general graphs.

In conclusion, adopting an empirical Bayes approach for estimating block memberships in a stochastic blockmodel, using an empirical prior obtained from a Gaussian mixture model estimate for the adjacency spectral embeddings,
can significantly improve block assignment performance.

\section*{Acknowledgments}\label{section:ack}
This work was supported in part by 
the National Security Science and Engineering Faculty Fellowship program, 
the Johns Hopkins University Human Language Technology Center of Excellence, 
the XDATA program of the Defense Advanced Research Projects Agency,
and the Erskine Fellowship program at the University of Canterbury, Christchurch, New Zealand.

\bibliographystyle{chicago}
\bibliography{EBref}
\end{document}